\pdfoutput=1
\documentclass[11pt]{article}
\usepackage[pdftex]{graphicx,color}
\usepackage{jheppub}
\usepackage{amsmath}
\usepackage{amssymb}
\usepackage{multirow}
\usepackage{mathtools}
\usepackage{bm}
\newcommand{\beq}{\begin{equation}}
\newcommand{\eeq}{\end{equation}}
\newcommand{\be}{\begin{equation}}
\newcommand{\ee}{\end{equation}}
\def \la{\langle}
\def \ra{\rangle}

\title{Boundary conformal field theory at the extraordinary transition:
       The layer susceptibility to $O(\varepsilon)$}
\author{M. A. Shpot}
\affiliation{Institute for Condensed Matter Physics, 79011 Lviv, Ukraine}
\emailAdd{shpot.mykola@gmail.com}

\abstract{We present an analytic calculation of the layer (parallel) susceptibility
at the extraordinary transition in a semi-infinite system with a flat boundary.
Using the method of integral transforms put forward by
McAvity and Osborn [Nucl. Phys. B {\bf 455}(1995) 522]
in the boundary CFT, we derive the coordinate-space representation of the
mean-field propagator at the transition point.
The simple algebraic structure of this function
provides a practical possibility of higher-order calculations.
Thus we calculate the explicit expression for the layer susceptibility
at the extraordinary transition in the one-loop approximation. Our result
is correct up to order $O(\varepsilon)$ of the $\varepsilon=4-d$ expansion and
holds for arbitrary width of the layer and its position in the half-space.
We discuss the general structure of our result and consider the limiting cases
related to the boundary operator expansion and (bulk) operator product expansion.
We compare our findings with previously known results and less complicated formulas in the case
of the ordinary transition. We believe that analytic results for layer susceptibilities
could be a good starting point for efficient calculations of two-point correlation functions.
This possibility would be of great importance given the recent breakthrough in bulk and
boundary conformal field theories in general dimensions.

\vspace{5mm}
}

\begin{document}
\maketitle

\section{Introduction}

The conformal invariance appeared in theoretical physics more than 100 years ago --- see \cite{FRW62,Kastrup08}. However, its importance for critical phenomena, associated with the second-order phase transitions has been recognized due to the seminal 1970 letter by Polyakov \cite{Polyakov70}.%
\footnote{The scale invariance inherent in systems with short-range interactions
at critical points, along with the invariance under translations and rotations, usually entails the existence of invariance under a larger symmetry group, that of conformal transformations \cite{Car87,DiFMS97,Hen99}.
The interrelations between the scale and conformal invariance are thoroughly discussed in the recent review paper \cite{Nakayama15}.}
About the same time, the conformal bootstrap equations came into play
\cite{Polyakov70,Migdal71,PP71},
along with the short-distance operator product expansions by Wilson \cite{Wilson69}
in the high-energy physics and Kadanoff \cite{Kadanoff69,KadanoffCeva71}
in the theory of critical phenomena.
The relevant relations have been the self-consistent equations for full
correlation functions of the Schwinger-Dyson type.
The term bootstrap implies an intrinsic self-consistency \cite{Chew68}.
These equations allowed conformally invariant solutions for correlation functions,
and the Operator Product Expansion (OPE) has been recognized as an efficient tool
to treat them.
An early review has been given by Mack \cite{Mack73}; a very recent review on Conformal
Field Theory (CFT) by the same author can be found in \cite{Mack19}.

In what followed, there has been much formal development, see
\cite{Parisi72,Sym72,FGG73,MackT73,Polyakov74,Mack77} and the reviews \cite{FP78,FP98},
but the practical output like explicit expressions or
numerical values of critical exponents appeared to be relatively modest
\cite{EPP71,PP72,Polyakov74} and \cite[Sec. 15]{FP78}.
In this respect, the greatest achievement of the classical conformal bootstrap programme
has been the calculations of critical exponents up to $O(N^3)$ in the large-$N$ expansion
\cite{VPH82,VDKS93,Gra94a,Gra94b,Gracey18}, see also the book by Vasiliev \cite{Vasiliev}
and the recent review \cite{Gracey18r}.
Moreover, the traditional bootstrap approach and OPE have been used in a comprehensive study of the operator algebra and multipoint correlation functions in a series of papers by Lang and R\"uhl, including \cite{LangR92}.

A breakthrough in the development of the CFT occurred with the appearance of the seminal paper by Belavin, Polyakov, and Zamolodchikov \cite{BPZ84} (see also \cite{BPZ84a} and \cite{ZZ89}).
In two space dimensions, where the group of conformal transformations is infinite-dimensional, the combination of the conformal bootstrap and the OPE was able to yield the exact solutions of the so-called minimal models.
The (infinite) family of these models included the well-known exactly solvable two-dimensional Ising, three-state Potts, and Ashkin-Teller models as special cases. It is hard to overestimate the impact of this work.
According to Google Scholar, by now the paper \cite{BPZ84} has more than 6000 citations, it has strongly influenced the Les Houches proceedings \cite{BZ90}, it was in the basis of the compilation \cite{ISZ98} reprinted there along with a series of related papers,
its methods and results have been discussed in numerous review articles and books, in particular \cite{Car87,FST89,DiFMS97,Hen99,Gaberdiel00}.
The two-dimensional CFT influenced several areas of mathematics and benefited from their development \cite{Gaberdiel00,Sch08,FRS10}.

The success of the CFT in two dimensions stimulated its application in different branches of
the Quantum Field Theory \cite{FradkinPalchik},
(super)strings \cite{Polch-1,Polch-2,GSW12-1,GSW12-2,Kaku}, statistical mechanics.
In the latter, essential progress has been achieved in
bulk \cite{OP94,Pet96} and
boundary critical phenomena \cite{Car84,MO93,MO95,ES94},
finite-size scaling \cite{CardyFSS},
critical Casimir effect \cite{FdG78,Krech,Car90,KG99,HHGDB08,BEK13}, 
polymer physics \cite{JK98,Eisenriegler}, etc. (see \cite{DiFMS97,Hen99}).
In view of the substantial progress of CFT, Henkel \cite{Hen02} proposed the
hypothesis of generalized local scale invariance intended to describe the dynamic
and strongly anisotropic static systems at criticality.
This conjecture and its implications in the
case of uniaxial Lifshitz points have been critically analysed in \cite{RDS11}.

A new explosion of interest in the conformal bootstrap approach has occurred in about the last
ten years, see the reviews \cite{Ps-D16,PRV19}. It started with the work \cite{RRTV08}
where new ideas and numerical methods have been applied to old equations.
The research activities moved to space dimensions higher than two, especially $d=3$.
The scaling dimensions of basic operators and hence the critical exponents have been
extracted from the fundamental consistency conditions and conformal symmetry considerations
without any microscopic input \cite{ElS12,ElS14,GR14,SD15,EHS16}.
The highlight of the "new" bootstrap approach is the most precise calculation
of critical exponents of the Ising model in $d=3$ dimensions \cite{Kos16}, namely
$(\eta,\;\nu)=(0.0362978(20), 0.629971(4))$.
As admitted by the authors of this last reference, their results for the $O(N)$ model with
$N=2$ and $3$ were still less accurate compared to the best Monte Carlo estimates \cite{Campostrini06,Campostrini02,HV11}.
Very recently, however, the new precise values for scaling dimensions
in the three-dimensional $O(2)$-symmetric model, compatible with Monte Carlo results,
have been obtained in \cite{Chester20}.%
\footnote{For an exhaustive comparison with
other data, especially that of the six-loop epsilon expansion, see also \cite{KomPa17}.}

Again, this development stimulated numerous research activities in diverse scientific fields.
Merely search "bootstrapping" in Google Scholar, and you will see hundreds of interesting contributions.
A significant number of references can be found in the review papers
\cite{Ps-D16,PRV19}.
Among them, of particular interest for us are recent investigations of systems with boundaries \cite[Sec. V.B.6]{PRV19}.\footnote{Exhaustive reviews on surface critical behavior in semi-infinite systems can be found in \cite{Bin83,Die86a,Die97}.}

The early fundamental work by Cardy "Conformal Invariance and Surface Critical Behavior"
\cite{Car84} appeared in 1984. The general formulation was given in $d$
dimensions, and some exact results were derived in $d=2$.
Approximately at the same time, the implications of the conformal invariance in
systems with boundaries have been noticed in several explicit calculations
(see the references in \cite[p. 82]{Car87}).
In semi-infinite two-dimensional CFTs,
Cardy and Lewellen \cite{CL91,Lewellen} introduced a system of "sewing constraints".
The essential idea, now ubiquitous in the modern conformal bootstrap approach,
was a self-consistent treatment of correlation functions whose
operators have been expanded at short distances both in the bulk \cite{Wilson69,Kadanoff69}
and near the boundary \cite{DD81c}.
A new essential portion of fundamental
knowledge has appeared in a series of papers by McAvity and Osborn
\cite{MO92}, \cite{MO93} and \cite{MO95}.
The last of these references provided a direct basis for
building up in \cite{LRR13} the modern bootstrap program for the boundary CFT in $d$ dimensions.

Liendo, Rastelli, and van Rees \cite{LRR13} formulated the self-consistent equations by comparing two different short-distance expansions for the same two-point correlation function.
These are the Operator Product Expansion and the Boundary-Operator Expansion (BOE) worked out in \cite{MO95} in a semi-infinite $d$ dimensional Euclidean space bounded by a flat $d-1$ dimensional (codimension-one) surface.
The equations could be checked analytically in two cases.
It was possible to reproduce the free Neumann and Dirichlet propagators in $d$ dimensions.
Also, the non-trivial $O(\varepsilon)$ contributions to the two-point functions in $d=4-\varepsilon$ dimensions could be derived for the special and ordinary transitions in accord with previous calculations \cite{GW85,MO93,MO95}.%
\footnote{The important relations between the boundary conditions and types of the
boundary critical behavior have been recently discussed in \cite{Die20}.}
In \cite[Appendix B]{LRR13} the mean-field
propagator of the extraordinary transition in $d=4$ has been reproduced (see \cite[(4.96)]{Jasnow} and Sec. \ref{EOC} below)%
\footnote{To the best of our knowledge, the $O(\varepsilon)$ correction to this function remained unknown until recently \cite{DHS20}.}.
The expression \cite[(B.37)]{LRR13} for the
one-point function $\langle\phi\rangle$ could be compared with the order-parameter profile of \cite{DS93}. Moreover, using the linear programming methods, the bootstrap constraints have been studied in $d=3$ for the special and extraordinary transitions.
This resulted in numerical estimates, some of which were compared with the results of two-loop calculations of \cite{DS98} (see also \cite{DS94,Sh97}).

Further progress in extracting the conformal data from the bootstrap equations in three-dimensional semi-infinite systems has been achieved in the subsequent papers \cite{GLMR15,Gliozzi16}.
Here the analysis of bootstrap equations differs from that of \cite{LRR13}.
The method of determinants is used, relying on a truncation of the operator spectrum.
In the case of the extraordinary transition, the bootstrap yielded certain information on the bulk operator spectrum \cite{GLMR15}, see also \cite{DHS20}.
For the ordinary and special transitions precise estimates have been obtained for relevant scaling dimensions, which compared well both with field-theoretic calculations \cite{DS98} and a series of Monte Carlo results: see \cite[Table 1]{Gliozzi16}.

Other types of "defects" such as line (codimesion-two) defects
as well as certain general features of the boundary CFT
have been studied in \cite{Billo13,Gai14,Billo16} and \cite{Herzog17,Rast2017,Mazac19,Kaviraj18};
see also the recent proceedings \cite{Andrei}.

Very recently, essential progress has been achieved in the analytic solution of boundary-CFT bootstrap equations \cite{BHS19}. First, the known anomalous scaling dimensions of the leading boundary operators corresponding to the special and ordinary transitions were reproduced
up to order $O(\varepsilon^2)$ \cite[p. 12]{BHS19}.
This is the highest order of the $\varepsilon$-expansion known up to now for critical exponents of the special and ordinary transitions.
Moreover, just "using the analytic structure of the two-point function and symmetries of the BCFT" the $O(\varepsilon^2)$ contribution to the scaling functions of the two-point correlators
have been found and written down in \cite[Sec. 4.4]{BHS19} (see also \cite[Sec. 2.4.1]{PS20})
without any comparison or further discussion. At order $O(\varepsilon^2)$,
these functions were not explicitly known in the statistical physics literature
(see, however, \cite[(B.45)-(B.46)]{DD81c}, \cite[(16)]{OO85}).
Moreover, the two-point correlation function of the extraordinary transition has been
derived to order $O(\varepsilon)$ only quite recently \cite{DHS20}.
In the rest of this paper we shall concentrate on this last type of transition.

The extraordinary transition has been introduced into the subject by Lubensky and
Rubin \cite{LR75b} in 1975.
The general information and relevant original literature on this transition
can be found in the standard references \cite{Bin83,Die86a,Die97}.
Its special feature is the presence of the non-vanishing order-parameter profile
$\langle\phi(z)\rangle$ both above and below the bulk critical temperature $T_c$
for $z\ge 0$, where $z$ is the coordinate normal to the hyperplane constraining the half-space
and located at $z=0$. All surface critical exponents (or corresponding scaling dimensions)
can be expressed in terms of the bulk exponents. The recognition of this feature goes
back to Bray and Moore \cite{BM77}
(see also \cite{BC87} where the conformal-invariance arguments are involved).
The most important feature from the conformal bootstrap perspective is that the
most relevant terms of the
boundary-operator expansion of a general bulk operator contain the identity operator
and the $zz$-component of the stress-energy tensor, $T_{zz}(0)$, with the scaling dimension $d$
\cite{Car90}, see also \cite[Sec. 3.3]{Die97}.

There are several interesting physical systems undergoing phase transitions which belong to the "extraordinary" universality class.
The most prominent example is the experimentally observed phenomenon of critical adsorption, see \cite{FdG78,FD95,Law01}.
It occurs close to the demixing point in liquid binary mixtures confined
by a hard wall or interface, and in constrained polymer solutions \cite{Eisenriegler}.

The theoretical description of the extraordinary transition is far from completeness.
To a large extent, this is due to mathematical difficulties one immediately encounters.
So, by presenting the new analytic results in this non-trivial case, we hope to stimulate some progress in this direction.

\vskip3mm
In Sec. \ref{PROP} we recall the known $pz$-representation of the mean-field propagator at the extraordinary transition.
Using one of the methods of the boundary conformal field theory we obtain the coordinate representation of this function at criticality and discover its simple and elegant structure.
In Sec. \ref{CHI} we present an explicit calculation of the layer susceptibility $\chi(z,z')$ for arbitrary distances $z$ and $z'$ from the surface in the one-loop approximation.
An explicit expression is obtained up to $O(\varepsilon)$ order of the $\varepsilon=4-d$ expansion.
The singularities it contains have their interpretation in terms of short-distance operator product expansions.
The results are compared with the previously known data and with similar expressions of the more common case of the ordinary transition.
In Sec. \ref{D:O} we discuss the relevance of present results for eventual
future calculations of coordinate representations of two-point correlation functions.
This is an interesting issue given the tremendous recent progress in the conformal field theory, both in the bulk and semi-infinite systems.
The Appendix contains some technical details.

\section{Mean-field correlation function at the extraordinary transition}\label{PROP}

We consider a scalar $\phi^4_d$ theory in a $d$-dimensional semi-infinite space
$\mathbb R^d_+=\{x=(\bm r,z)\in\mathbb R^d\mid\bm r\in\mathbb R^{d-1},z\ge0\}$.
The coordinate $z\ge0$ is orthogonal to a flat $(d-1)$-dimensional
boundary located at $z=0$. Momenta $\bm p$ are conjugated to
position vectors $\bm r$ in hyperplanes parallel to the boundary.

At the extraordinary transition%
\footnote{Exhaustive background information on the underlying model and phase transitions
can be found in \cite{Die86a,DS93,ES94}. For a short recent introduction see \cite{DHS20}.},
the mean-field propagator in the "mixed" $pz$ representation is given by \cite[(2.5)]{ES94}
\beq\label{pzprop}
G_0(p;z<z')={1\over 2 p}\,\Big[W(-pz)-W(pz)\Big]\,W(pz')
\eeq
where the function $W$ is
\beq\label{w}
W(x)={\rm e}^{-x} \left(1+{3\over x}+{3\over x^2}\right)\,.
\eeq
The function $G_0(p;z,z')$ obeys the equation \cite[Sec. II]{Eisenriegler84},
cf. \cite[App. A]{ES94}, \cite[Sec. II]{LR75b}
\beq\label{GRE}
\left[-\frac{d^2}{dz^2}+p^2+\frac{u_0}2\,m_0^2(z)\right]G_0(p;z,z')=\delta(z-z')
\eeq
with the boundary condition on the surface
\be\label{BCG}
\left(-\frac{d}{dz}+c_0\right)\left.G_0(p;z,z')\right|_{z=0}=0\,,\quad z'>0\,.
\ee
In \eqref{GRE}-\eqref{BCG}, $u_0$, the $\phi^4$ coupling constant,
and $c_0$, the surface enhancement, are the parameters of the underlying
Landau-Ginzburg Hamiltonian (see e. g. \cite{DHS20}).
We set the "mass" $\tau_0$ to zero, as we are interested only in the theory
right at the transition point $T=T_c$.

In \eqref{GRE}, $m_0(z)$ is the critical order-parameter
profile in the Landau (mean-field) approximation.
It is the solution of the "classical equation of motion"
\be\nonumber
-\frac{d^2}{dz^2}\,m_0(z)+\frac{u_0}6\,m_0^3(z)=0
\ee
with the boundary condition
\be\nonumber
\left(\frac{d}{dz}-c_0\right)m_0(z)|_{z=0}=-h_1\,,
\ee
where $h_1$, the local external ordering field at the surface comes into play.
When $T=T_c$, the profile $m_0(z)$ vanishes when approaching the bulk limit at $z\to\infty$.
The mean-field results \eqref{pzprop}-\eqref{w} for $G_0(p;z,z')$ and
$m_0(z)=\sqrt{12/u_0}\;z^{-1}$ are the same in two limits: $c_0\to-\infty$ with $h_1=0$,
and $h_1\to\infty$ with any $c_0$.%
\footnote{Though producing the same results, these two limits have different physical
background: see footnote 2 in \cite{DHS20} and references therein; we shall not distinguish
between them and continue to call the transition "extraordinary" anyway.}

The propagator $G_0(p;z,z')$ is that of Gaussian fluctuations of the order parameter around the
mean-field solution $m_0(z)$. It is the \emph{connected} two-point function
in the (unperturbed) zero-loop, or tree approximation of the Landau-Ginzburg theory.
This is the lowest-level approximation in the famous Feynman-graph expansion in
the number of loops.
As in \cite{LR75b}, $G_0(p;z,z')$ is often called the mean-field propagator.

The propagator \eqref{pzprop}-\eqref{w} belongs to the theory with a symmetry breaking.
Just as in the more common case of the bulk theory at $T<T_c$, the presence of the term
$u_0\phi^4$ in the Landau-Ginzburg Hamiltonian is necessary for an adequate
description of the ordered phase already in the lowest approximation.
This term cannot be simply neglected as it is done in the usual "free" theory.

It is well known that the mean-field theory is correct in $d\ge4$.
When $d<4$, the loop corrections are important, and near $T_c$ the mean-field results
may only be seen as certain approximations to the correct non-classical ones.%
\footnote{At the extraordinary transition,
the one-loop corrections to the zero-loop propagator $G_0(r;z,z')$
have been recently derived in \cite{DHS20}.}
In the absence of the external ordering field on the boundary,
for supercritical enhancement of surface interactions $c_0<0$,
the extraordinary transition with a one-component order parameter exists for all $d>3$.
Its mean-field treatment in generic dimensions $d$ has been carried out
in the pioneering work \cite{LR75b}.
The statement of \cite[p. 12]{LRR13} that
"the extraordinary transition\ldots appears at first order in the Wilson-Fisher fixed point
in $d=4-\varepsilon$ dimensions" is incorrect.

Equally incorrect is the claim of \cite[p. 43]{LRR13}
that "there is no extraordinary transition in $4$ dimensions\ldots".
As we have seen above, there is nothing to prevent the mean-field theory of the
extraordinary transition from existing in four dimensions provided $u_0\ne0$.
The mean-field correlation function in $d=4$ has been derived in \cite[p. 3900]{LR75b},
see also \eqref{gd4} and related references.
This function contains a logarithmic term, which arises in a non-trivial limiting process
when $d\to4$.

Indeed, the marginal case of the upper critical dimension $d^*=4$ is special.
In four dimensions, the theory is asymptotically free, and the $\phi^4$ coupling constant
is a marginally irrelevant variable.
There is the IR-stable fixed point $u^*=0$ in this theory, but one has to keep in mind that
the running coupling constant $u(l)$,
which replaces $u_0$ in renormalization-group flow equations,
behaves as $u(l)\sim|\ln l|^{-1}$ when the flow parameter $l\to0$.
This generally leads to \emph{logarithmic corrections}
to the mean-field behavior in $d=4$ \cite[Sec. VIII.B]{BLZ76}, \cite[Sec. 9-6]{Amit},
\cite[Sec. 5.6]{Car96a}.
In the context of semi-infinite systems, these corrections are discussed in
\cite[Sec. III.C.14]{Die86a}.
The one-point function $m(z)$ at criticality contains the factor $1/\sqrt{u(l)}$.
With the appropriate choice of the flow parameter $l=(\mu z)^{-1}$, where $\mu$
is an arbitrary momentum scale, its presence leads to the
form $m(z)\sim z^{-1}\sqrt{|\ln\mu z|}$ in $d=4$%
\footnote{H. W. Diehl, private communication}.
This is instead of an uncontrolled divergence which would appear in an illegitimate
attempt to use $u_0=0$ or $u^*\sim\varepsilon=0$ when considering $m(z)$ at $d=4$.

In the following subsections we shall derive $G_0(r;z,z')$,
the counterpart of \eqref{pzprop} in the coordinate representation.
The function $G_0(r;z,z')$ can be found from \eqref{pzprop} by a direct Fourier transformation
of $G_0(p;z,z')$ in $d-1$ directions parallel to the boundary.
However, we find this way a bit annoying and not very instructive.
Instead of this, we shall use another integral transformation that employs
the conformal invariance of the underlying theory.
It was introduced to the subject in \cite{MO95} and \cite{McA95}. For convenience, we reproduce
the needed information in a few words below.

\subsection{The integral transformation}
In a semi-infinite critical system, the conformal invariance
implies that the two-point correlation function of bulk fields can be written
in the scaling form
\beq\label{scalg}
G(x,x')=\frac1{(4zz')^{\Delta_\phi}}\;g(\xi)\,.
\eeq
Here $\Delta_\phi$ is the usual full scaling dimension of the field%
\footnote{$\eta$ is the familiar (bulk) correlation function exponent,
and $\eta/2$ is the anomalous dimension of the field. The $\varepsilon$-expansion
of $\eta$ starts with $O(\varepsilon^2)$.}%
$\Delta_\phi=\frac12(d-2+\eta)$ and the variable $\xi$ is one of the possible
invariants of the (restricted) conformal transformations preserving the boundary
of the system \cite{Car84}, \cite{Car87}:
\beq\label{xid}
\xi= {|x-x'|^2\over 4zz'}\,.
\eeq
Integration of the correlation function $G(x,x')$ over directions parallel
to the surface in a strip between the parallel hyperplanes at $z$ and $z'$
defines the layer (or parallel) susceptibility
\beq\label{xigen}
\chi(z,z')=\int d^{d-1}r\,G(r;z,z')\,.
\eeq
Here $r$ is the distance between the points $x$ and $x'$ in
parallel directions: the vector $x'-x$ has components $({\bm r},z'-z)$.
Using this convention in the definition of $\xi$
in (\ref{xid}) and reporting
the scaling form (\ref{scalg}) into (\ref{xigen}) we obtain,
after performing the angular integration,
\beq\label{xang}
\chi(z,z')=\frac1{(4zz')^{\Delta_\phi}}\,S_{d-1}
\int_0^\infty\!\! dr\,r^{d-2} \,g\! \left({r^2{+} |z{-}z'|^2\over 4zz'}\right)\,.
\eeq
Here $S_{d-1}$ is the area of a unit sphere embedded in $\mathbb R^{d-1}$. For general $d$,
the well-known formula is
\beq\label{Sd}
S_d={2 \pi^{d\over 2}\over \Gamma\left({d\over 2}\right)}\,.
\eeq
A slight modification of the integration variable in (\ref{xang})
leads to the representation
\beq\label{interm}
\chi(z,z')={1\over (4zz')^{\Delta_\phi-\frac{d-1}{2}}}\,{S_{d-1}\over 2}
\int_0^\infty\!\! du\; u^{d-3\over 2} \,g(u{+}\rho)
\eeq
which defines the general scaling form for the layer susceptibility (cf. (\ref{SX}) below)
\beq\label{xisc}
\chi(z,z')=(4zz')^{1-\eta\over 2}\;\hat g(\rho)\,.
\eeq
In the last two expressions $\rho$ denotes the combination
\beq\label{ro}
\rho={|z-z'|^2\over 4zz'}
\eeq
explicitly appearing in (\ref{xang}).

An important feature is that the equations (\ref{interm}) and (\ref{xisc})
define an integral transform
\beq\label{gro}
\hat g(\rho)={\pi^\lambda\over \Gamma(\lambda)}
\int_0^\infty\! du\; u^{-1+\lambda} \,g(u+\rho)
\qquad\mbox{with}\qquad \lambda=\frac{d-1}{2}
\eeq
which can be inverted via
\beq\label{inv}
g(\xi)={\pi^{-\lambda}\over\Gamma(-\lambda)}
\int_0^\infty\! d\rho\; \rho^{-1-\lambda} \,\hat g(\rho+\xi)\,.
\eeq
The inversion formula \cite{MO95,McA95} allows to calculate the scaling function $g(\xi)$
of the correlator $G(x,x')$ starting from the scaling function $\hat g(\rho)$ of the
layer susceptibility $\chi(z,z')$ which is normally easier accessible
in the explicit (perturbative) calculations, see \cite[Sec.3.2]{DHS20}.

\subsection{A warm-up: Ordinary transition}
Let us demonstrate the above procedure with a simplest example by reproducing
the well-known Dirichlet propagator $G_\mathrm{D}(x,x')$.
The massless Dirichlet propagator in the $pz$ representation $G_\mathrm{D}(p;z<z')$
is recovered from $G_0(p;z<z')$ in (\ref{pzprop})
when we take there into account only the pure exponential terms in the functions $W$.
The limit $p\to 0$ of $G_\mathrm{D}(p;z<z')$ leads us to the layer susceptibility
$\chi_\mathrm{D}(z<z')=z$. In a more general, symmetric form
\be\nonumber
\chi_\mathrm{D}(z,z')=\mbox{min}(z,z')=\frac{z+z'-|z'-z|}2\,.
\ee
This result should be represented in the scaling form (\ref{xisc}). We have
\be\nonumber
\chi_\mathrm{D}(z,z')=\sqrt{4zz'}\; \frac12\!
\left({\mbox{min}(z,z')\over \mbox{max}(z,z')}\right) ^\frac12
\equiv \sqrt{4zz'}\; \frac12\;\zeta^\frac12\,.
\ee
Here we introduced a new variable $\zeta$,
\be\nonumber
\zeta={\mbox{min}(z,z')\over \mbox{max}(z,z')}\,.
\ee
It will play an important role in subsequent one-loop calculations. But for the
present purpose we need to identify the factor $\frac12\,\zeta^\frac12$
as the scaling function $\hat g_\mathrm{D}(\rho)$ depending on the scaling variable $\rho$.
In terms of $\zeta$, the definition (\ref{ro}) of $\rho$ reads
\be\nonumber
\rho={1\over 4}\,(\zeta+\zeta^{-1}-2)\,,
\ee
and thereof follows
\be\nonumber
\zeta=(\sqrt{\rho+1}-\sqrt{\rho})^2\,.
\ee
Hence we obtain for the required scaling function
\be\nonumber
\hat g_\mathrm{D}(\rho)=\frac12\left(\sqrt{\rho+1}-\sqrt{\rho} \right)\,.
\ee
Thus, in order to calculate $g_\mathrm{D}(\xi)$ through (\ref{inv}) we should evaluate the
Euler integral in
\beq\label{eul}
g_\mathrm{bulk}(\xi)=-\,{\pi^{-\lambda}\over 2\Gamma(-\lambda)}
\int_0^\infty\! d\rho\; \rho^{-\lambda-1} \,\sqrt{\rho+\xi}
\eeq
and subtract the same function $g_\mathrm{bulk}$ of the argument $\xi+1$. The result is
\beq\label{gdsc}
g_\mathrm{D}(\xi)=C_d\left[\xi^{1-{d\over 2}}-(\xi+1)^{1-{d\over 2}}\right]\,.
\eeq
Multiplying this scaling function through the overall factor $(4zz')^{-\Delta_\phi}$
according to (\ref{scalg})%
\footnote{Remembering that in the present zero-loop approximation $\eta=0$.}
and using the explicit expression (\ref{xid}) for the
scaling variable $\xi$ we obtain the Dirichlet propagator in its usual form given
in terms of the Euclidean coordinates in $\mathbb R^d$:
\beq\label{dprop}
G_\mathrm{D}^{(d)}(r;z,z')=C_d\,\left(R_{-}^{2-d}-R_{+}^{2-d} \right)
=G^{(d)}_\mathrm{bulk}(r;|z-z'|)-G^{(d)}_\mathrm{bulk}(r;z+z')\,.
\eeq
It consists of the difference of two usual free bulk propagators given by
$G^{(d)}_\mathrm{bulk}(R)=C_d\,R^{2-d}$ and
depending respectively on two distances $R_-$ and $R_+$ in $\mathbb R^d$:
\beq\label{Rmp}
R_{\mp}=\sqrt{r^2+(z \mp z')^2},\;\qquad r^2=\sum_{\alpha=1}^{d-1}r_\alpha^2.
\eeq
The presence of the bulk functions in \eqref{dprop} explains the appearance of the
subscript "bulk" in the function $g_\mathrm{bulk}$ in \eqref{eul}
The constant $C_d$ stemming from \eqref{eul} is
\beq\label{cd}
C_d={1\over 4}\pi^{-d/2}\Gamma\Big({d-2\over 2}\Big)={S_d^{-1}\over d-2}
\eeq
with $S_d$ from \eqref{Sd}.
This is the usual "geometric factor" appearing in the coordinate representation
of free propagators (cf. \cite[(3.3)]{MO95}).

\subsection{The "extraordinary" case}\label{EOC}
Once we have carefully done the above exercise, there remains not much work to derive a similar coordinate representation of the mean-field propagator at the extraordinary transition.
This will correspond to the Fourier transformation over parallel directions of the "$pz$" propagator in (\ref{pzprop}).
Again, the $p\to 0$ limit of this expression yields the mean-field layer susceptibility
\beq\label{mfs}
\chi_0(z,z')=\frac15 {\left[\mbox{min}(z,z')\right]^3\over \left[\mbox{max}(z,z')\right]^2}
=\sqrt{4zz'}\;{1\over 10}\;\zeta^{5\over 2}\,.
\eeq
Hence, the scaling function to be integrated in (\ref{inv}) is
\beq\label{sfti}
\hat g(\rho+\xi)={1\over 10}\left(\sqrt{\rho+\xi+1}-\sqrt{\rho+\xi} \right)^5\,.
\eeq
The integral is not as complicated as it might appear at first glance.
Simple algebraic properties
\beq\label{alg1}
\left(\sqrt{\rho+\xi+1}-\sqrt{\rho+\xi}\right)\left(\sqrt{\rho+\xi+1}+\sqrt{\rho+\xi}\right)=1
\eeq
and
\beq\label{alg2}
\left(\sqrt{\rho+\xi+1}-\sqrt{\rho+\xi}\right)^2
+\left(\sqrt{\rho+\xi+1}+\sqrt{\rho+\xi}\right)^2=2\,(1+2\rho+2\xi)
\eeq
reduce $\hat g(\rho+\xi)$ to a linear combination of square roots
thus leading us to several simple integrals similar to that in (\ref{eul}).
We write $a^5$ in (\ref{sfti}) as $a\,a^2a^2$, express the first $a^2$ through (\ref{alg2})
and eliminate the resulting "mixed" terms using (\ref{alg1}).
Then we perform the same trick with the remaining factor $a^2$. The result is
\begin{align}\label{i5}
\left(\sqrt{\rho+\xi+1}-\sqrt{\rho+\xi} \right)^5&=
\left[-5-20(\rho+\xi)-16(\rho+\xi)^2\right]\sqrt{\rho+\xi}\\
&+\left[5-20(\rho+\xi+1)+16(\rho+\xi+1)^2\right]\sqrt{\rho+\xi+1}\,.\nonumber
\end{align}
We see that taking only the first terms $-5$ and $+5$  in square brackets
again reproduces the "Dirichlet" contribution (\ref{gdsc}) in the present
scaling function. Performing explicitly the remaining integrations related to
the first line in (\ref{i5}) we obtain
\be\nonumber
g^{(1)}(\xi)=C_d\,\xi^{1-{d\over 2}}-{6\over \pi}\,C_{d-2}\,\xi^{2-{d\over 2}}
+{12\over \pi^2}\,C_{d-4}\,\xi^{3-{d\over 2}}\,.
\ee
The second line of (\ref{i5}) produces, up to the signs, the contributions
of the same form but with replacements $\xi\to (\xi+1)$. Adding up all terms we
obtain for the whole scaling function%
\footnote{This agrees with the result \cite[(C11)-(C12)]{LR75b} at $\lambda=0$.}
\beq\label{gxi}
g(\xi){=}C_d\!\left[\xi^{1-{d\over 2}}-(\xi+1)^{1-{d\over 2}}\right]
-{6\over \pi}C_{d-2}\! \left[\xi^{2-{d\over 2}}+(\xi+1)^{2-{d\over 2}}\right]+
{12\over \pi^2}C_{d-4}\!\left[\xi^{3-{d\over 2}}-(\xi+1)^{3-{d\over 2}}\right].
\eeq
The increase of powers of $\xi$ by one can be traced back to a decrease of $d$ by two.
This is anticipated in the coefficients $C_d$, $C_{d-2}$, and $C_{d-4}$ given by (\ref{cd}).
Thus we may write
\be\nonumber
g(\xi)=g_\mathrm{D}^{(d)}(\xi)-{6\over \pi}\,g_\mathrm{N}^{(d-2)}(\xi)
+{12\over \pi^2}\,g_\mathrm{D}^{(d-4)}(\xi)\,.
\ee
Here $g_\mathrm{D}^{(d)}(\xi)$ is the scaling function of the Dirichlet propagator
from (\ref{gdsc}).
Similar functions of the same argument $g_\mathrm{N}^{(d-2)}$ and $g_\mathrm{D}^{(d-4)}$
have the functional form of the
Neumann propagator in $d-2$ and Dirichlet propagator in $d-4$ dimensions.
Along with the factor $(4zz')^{-\Delta_\phi}$ we can write
\beq\label{prop}
G_0^{(d)}(r;z,z')=G_\mathrm{D}^{(d)}(r;z,z')-
\frac{3}{2\pi}\frac{1}{z z'} G_N^{(d-2)}(r;z,z')+
\frac{3}{(2\pi)^2}\frac{1}{(z z')^2}G_\mathrm{D}^{(d-4)}(r;z,z').
\eeq
Though the "Neumann" and "Dirichlet" functions in \eqref{prop}
are parametrized by $d-2$ and $d-4$, they still depend on distances $R_-$ and $R_+$ in $\mathbb R^d$ (see \ref{Rmp}). Generally,
\beq\label{gd'}
G_{D/N}^{(d')}(r;z,z')=G^{(d')}_\mathrm{bulk}(r;|z-z'|)\mp G^{(d')}_\mathrm{bulk}(r; z+z')
=C_{d'}\; (R_{-}^{2-d'} \mp R_{+}^{2-d'} ) \ .
\eeq
The $\varepsilon\to0$ limit of \eqref{prop} correctly yields%
\footnote{
This is compatible with \cite[(B.35)]{LRR13}, \cite[(C15)]{LR75b}
and  \cite[(4.96)]{Jasnow}
(equations \cite[(12)]{RJ82} and \cite[(5.48)]{Jas84} are incorrect).}
\beq\label{gd4}
G_0^{(d=4)}(x,x')=\frac{1}{4 z z'}\frac{1}{4\pi^2}
\left[\frac{1}{\xi}-\frac{1}{\xi+1}+12+6(1+2\xi)\ln{\xi\over \xi+1}\right]\,.
\eeq
The first term inside the square brackets is related to the bulk propagator in four dimensions, the first two terms correspond to the Dirichlet propagator, and the last two contributions give the correction specific for the present case of the extraordinary transition.
The logarithmic term arises in the limit $\varepsilon\to0$ due to simple $1/\varepsilon$ poles in
the coefficients $C_{d-2}$ and $C_{d-4}$ in (\ref{gxi}).
Similar logarithms appeared \cite{MO95} in a large-$N$ expansion of a
correlation function in the non-linear $O(N)$ $\sigma$-model, and their origin was discussed using the OPE arguments.
It seems that the mechanisms of the occurrence of logarithmic terms in both cases, as well as their interpretations from the point of view of
short-distance expansions could be similar.

The inverse $(d-1)$-dimensional Fourier transformation of (\ref{prop})
to the {\it pz} representation leads to an expression in terms of modified
Bessel functions $K_\nu(x)$ \cite{AS}
\begin{align}\label{pzk}
G_0(p;z,z')&={1\over\sqrt{2\pi}}\! \left[
\left({|z_-| \over p} \right)^{\frac12}\!\! K_{\frac12}(p |z_-|)
-{3 \over z z'}\! \left({|z_-| \over p} \right)^{{3\over 2}}\!\! K_{{3\over 2}}(p |z_-|)
+{3 \over (z z')^2}\! \left({|z_-| \over p} \right)^{{5\over 2}}\!\! K_{{5\over 2}}(p |z_-|)\right]
\nonumber\\
&-{1\over\sqrt{2\pi}}\! \left[
\left({z_+ \over p} \right)^{\frac12}\! K_{\frac12}(p z_+)
+{3 \over z z'} \left({z_+ \over p} \right)^{{3\over 2}}\! K_{{3\over 2}}(p z_+)
+{3 \over(z z')^2} \left({z_+\over p} \right)^{{5\over 2}}\! K_{{5\over 2}}(p z_+)\right]
\end{align}
where $z_-\equiv z'-z$ and $z_+\equiv z+z'$.
It is fully symmetric in its arguments, has clear structure and
actually reduces, after some algebra, to the familiar form (\ref{pzprop}) owing to the
simplifications in Bessel functions of half-integer orders \cite{AS}.
This result follows from a single transformation formula of
functions $G^{(d')}_\mathrm{bulk}$ from (\ref{gd'})
\be\nonumber
G^{(d')}_\mathrm{bulk}(p;z){=}\!\!\int\!\! d^{d-1}r{\rm e}^{i\bm p \bm r}G^{(d')}_\mathrm{bulk}(r;z)
{=}(2\pi)^{\nu+1}\,p^{-\nu}
\!\!\!\int_0^\infty\!\! dr\,r^{\nu+1} J_\nu(pr)\,G^{(d')}_\mathrm{bulk}(r;z)
{=}{1\over 2\pi} \Big({2\pi z \over p} \Big)^\delta\! K_\delta(p z)
\ee
with $\nu=(d-1)/2-1$ and $\delta=(d-d'+1)/2$. The associated inverse transformation is
\be\nonumber
G^{(d')}_\mathrm{bulk}(r;z)=\int\!\!{d^{d-1}p\over (2\pi)^{d-1}}\;{\rm e}^{-i\bm p \bm r}\,
G^{(d')}_\mathrm{bulk}(p;z)
=(2\pi)^{-(\nu+1)}\,r^{-\nu}\!\!\int_0^\infty dp\,p^{\nu+1} J_\nu(pr)\,G^{(d')}_\mathrm{bulk}(p;z).
\ee
Here $J_\nu(x)$ is the Bessel function of the first kind \cite{AS},
the required integrals can be found in standard mathematical references \cite{GR,PBM2}.

Obviously, the coordinate representation (\ref{prop}) is considerably simpler than (\ref{pzprop})
or (\ref{pzk}) and can be much more advantageous
for eventual higher-order calculations. In the next section we employ it
in a one-loop calculation of the layer susceptibility.

\section{The layer susceptibility}\label{CHI}
In this section we discuss an explicit calculation of the layer susceptibility
$\chi(z,z')$ at the extraordinary transition to the one-loop order.

\subsection{Perturbation theory}\label{PCHI}

The susceptibility of a layer confined between the planes $z=z$ and $z=z'$
(suppose $z\le z'$) is defined as the integral of the (connected) two-point correlation
function $G(r;z,z')$ with respect to the parallel coordinates $\bm r$, see
\cite{RJ82a,ES94,MO95} and \cite[Sec. III.B]{KlD99}.
It coincides with the $p=0$ limit of the correlation function in its {\it pz} representation:
\beq\label{xig2}
\chi(z,z')=\int d^{d-1}r\,G(r;z,z')=G(p=0;z,z')\,.
\eeq

Thus, the first-order Feynman-diagram expansion for the layer susceptibility is
given in the graphical form by%
\footnote{For more details on the loop expansion see \cite{ES94} and \cite{DHS20};
for convenience, we shall follow here the notation of \cite{ES94} very closely.
}
\beq\label{graphs}
\chi(z,z')=\raisebox{-15pt}{\includegraphics[width=31pt]{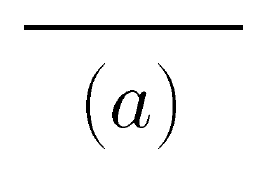}}
+\raisebox{-12pt}{\includegraphics[width=31pt]{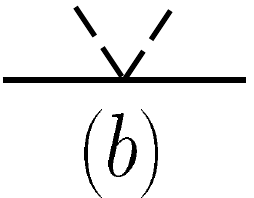}}
+\raisebox{-12pt}{\includegraphics[width=16pt]{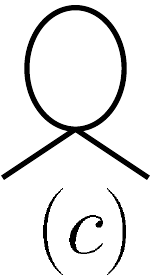}}
+\raisebox{-13pt}{\includegraphics[width=31pt]{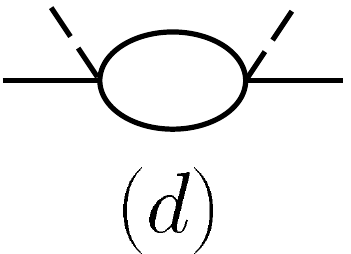}}\;.
\eeq
The first contribution (a) is the mean-field layer susceptibility \eqref{mfs}
\begin{align}\label{chi0}
\chi_a(z,z')=\chi_0(z,z')=G_0(p=0;z,z')&=
\frac15 {\left[\mbox{min}(z,z')\right]^3\over \left[\mbox{max}(z,z')\right]^2}=
{1\over 10}{(z+z'-|z-z'|)^3\over (z+z'+|z-z'|)^2}=\nonumber\\
&=\frac15 \,{z^3\over z'^2} \;\;\;\;\; \quad\mbox{if}\quad z<z' \;.
\end{align}

The "one-loop" terms are
\beq\label{BEB}
\chi_b(z,z')=-u_0 \int_0^\infty \! dy\,G_0(p=0;z,y)G_0(p=0;y,z') m_0(y) m_1(y)\;,
\eeq
\beq\label{chic}
\chi_c(z,z')=-{u_0\over 2}\int_0^\infty\!dy\,G_0(p=0;z,y) G_0(p=0;y,z')G_0(r=0;y,y)\,,
\eeq
and
\beq\label{d}
\chi_d(z,z')={u_0^2\over 2}\!\int_0^\infty\!\!dy\int_0^\infty\!\! dy'\,
G_0(p{=}0;z,y) m_0(y)\left[\int\!\!d{\bm r}\,G_0^2(r;y,y')\right]m_0(y')G_0(p{=}0;y',z')\,.
\eeq
A great simplifying feature here is that the propagators at external
lines are simply the corresponding mean-field layer susceptibilities (\ref{chi0}).

The vertices are associated with the bare coupling constant $u_0$ of the underlying
scalar $\phi^4$ theory, while
$m_0(z)$ and $m_1(z)$ are the zero- and one-loop contributions of the magnetization profile at the transition point:
\beq\label{prof}
m(z)=m_0(z)+m_1(z)=\sqrt{12\over u_0}\,{1\over z}
-{\sqrt{3 u_0}\over (1{+}\varepsilon) (4{-}\varepsilon)}\; z\; G_0(r{=}0;z,z) \;.
\eeq
In the dimensional regularization which implies the prescription
$G_\mathrm{bulk}(R=0)=0$ (see (\ref{dprop}) -- (\ref{cd}))
the self-energy "tadpole" $G_0(r{=}0;z,z)$ is given by
\beq\label{loop}
G_0(r{=}0;z,z)= \int\frac{d^{d-1}p}{(2\pi)^{d-1}}\;G_0(p;z,z)={\cal P} \,z^{-2+\varepsilon}
\eeq
with the constant
\be\nonumber
{\cal P}={\cal P}_D {(4{-}\varepsilon)(6{-}\varepsilon)\over 2{+}\varepsilon}\,{1\over \varepsilon}\;.
\ee
Here, the factor
\beq\label{PD}
{\cal P}_D=- \,C_d\,2^{-2+\varepsilon} \qquad\mbox{with}\qquad  C_d={S_d^{-1}\over d-2}
\eeq
just coincides with the analogous proportionality
constant ${\cal P}_D$ from the tadpole graph including the Dirichlet propagator (\ref{dprop}):
\beq\label{DT}
G_D(r{=}0;z,z)= \int\frac{d^{d-1}p}{(2\pi)^{d-1}}\;G_D(p;z,z)={\cal P}_D \,z^{-2+\varepsilon}\;.
\eeq

In fact, at this stage we have initiated the renormalization of the layer susceptibility
\eqref{xig2}-\eqref{graphs}. The elimination of the bulk tadpole in the diagram (c)
corresponds to the mass renormalization (see e. g. \cite{Amit}) in the present
"massless" theory, which actually corresponds to a fine-tuning of the temperature
to the bulk critical temperature $T_c$.
In doing this, we find the coefficient at $(2z)^{-2+\varepsilon}$ in \eqref{loop} via
\be\nonumber
\lim_{\xi\to0}\left[g(\xi)-C_d\xi^{1-\frac d2}\right].
\ee
This means that we are looking for the finite part of the scaling function $g(\xi)$ from
\eqref{gxi} in the asymptotic limit $\xi\to0$.
Application of the same procedure just to the first term in \eqref{gxi} obviously leads to
the result \eqref{PD} for the Dirichlet tadpole \eqref{DT}.

An essential difference between the one-loop tadpole graphs in the cases of the extraordinary and ordinary transitions is that ${\cal P}$ contains a pole in $\varepsilon$, while ${\cal P}_D$ is of order $O(\varepsilon^0)$.
The $1/\varepsilon$ pole in ${\cal P}$ stems from the coefficient functions $C_{d-2}\sim\Gamma(d/2-2)$ and $C_{d-4}\sim\Gamma(d/2-3)$
in the second and third terms of (2.34), which are absent in the Dirichlet case.
This dimensional dependence is specific for the present theory where
the presence of a non-zero order parameter profile at $T=T_c$ influences the zero-loop
two-point function given by $g(\xi)$ in \eqref{gxi}.

The appearance of the $1/\varepsilon$ pole in the diagram (c) and similarly in other one-loop
graphs to be considered further,
prevents from calculating their individual contributions at $d=4$ as it is usually possible in less complicated theories.
The renormalization usually takes care of divergent diagrams
providing subtractions that render them finite at $d=4$.
This is very nicely described in the book by Amit \cite{Amit}.
In the present $\varepsilon$ expansion of the layer susceptibility,
one has to do the one-loop integrals by keeping finite $\varepsilon>0$ and producing
their Laurent expansions up to $O(1)$ for each individual graph.
The final cancelation of singular terms occurs between the different graphs in the entire
one-loop contribution.%
\footnote{It was, however, technically convenient and aesthetically appealing
to compute each Feynman integral
in generic $d=4-\varepsilon$ dimensions and find the finite $\varepsilon\to0$ limit of the whole
combination of one-loop graphs as in \eqref{INS}.}

The necessity of such \emph{mutual} cancelation of $1/\varepsilon$ poles in one loop
will become immediately clear when we finish the discussion of the renormalization
of the layer susceptibility, which we have already started above.
There is no alternative source of cancelation of one-loop $1/\varepsilon$ poles due to renormalization
process.

In fact, the layer susceptibility is just a correlation function
of two fields away from the boundary, taken at zero parallel momentum.
Thus, the renormalization of this function proceeds just in the same
way as in an infinite bulk (a very careful and detailed discussion of all possible
correlation functions in semi-infinite systems can be found in \cite{Die86a}).
As is well-known \cite{Amit}, apart from the mass renormalization discussed just above,
this implies the field renormalization.
The involved renormalization factor $Z_\phi$ comes into play only at the
two-loop level, and in the present one-loop calculation, we have trivially $Z_\phi=1$.
Finally, we have to do the vertex renormalization.
However, the bare $\phi^4$ coupling constant $u_0$ appears only in the
one-loop contribution. Its renormalization is thus again trivial, given in \eqref{u*}.
As in the case of $Z_\phi$, the non-trivial vertex renormalization factor $Z_u$ would remove
some $1/\varepsilon$ poles only starting from two loops.%
\footnote{
By contrast, in the one-loop calculation of the renormalized order-parameter profile $m(z)$
in \eqref{prof}, the $1/\varepsilon$ pole of the one-loop term gets canceled by
the renormalization of $u_0$ in the leading mean-field contribution: see \cite{DS93},
\cite[App. A]{DHS20}. Note that the exponent $\beta/\nu$ used throughout \cite{DS93}
is just the scaling dimension of the field: $\beta/\nu=(d-2+\eta)/2=\Delta_\phi$.
This is a direct consequence of usual (hyper)scaling relations.
}

Let us return to the actual calculations.
The one-loop graphs (b) and (c) contain the same Feynman integrals owing to
the definitions of $m_0(z)$ and $m_1(z)$ in (\ref{prof}). They are easy to compute,
and their sum is explicitly given in \cite[(4.6)]{ES94}:
\beq\label{b+c}
\chi_b(z,z')+\chi_c(z,z')=\chi_0(z,z') \,{u_0\over 10}\,{\cal P}
\left[{12\over (1{+}\varepsilon)(4{-}\varepsilon)} -1\right] z'^\varepsilon \!
\left({\zeta^\varepsilon \over 5+\varepsilon}+{1{-}\zeta^\varepsilon \over \varepsilon}+{1\over 5-\varepsilon} \right).
\eeq
Here again,
\beq\label{zeta}
\zeta={\mbox{min}(z,z')\over \mbox{max}(z,z')}=
{z+z'-|z-z'|\over z+z'+|z-z'|}
\eeq
and with the convention $z\le z'$, we have $\zeta=z/ z'\le 1$.

The calculation of the graph (d)
is much more involved, and it was performed in \cite{ES94} only in the asymptotic regime
$\zeta\ll 1$.
Nevertheless, using our real-space zero-loop propagator (\ref{prop})
we were able to find an explicit expression for its contribution.
The most complicated issue was the calculation of the inner loop of the graph (d)
in the square brackets of \eqref{d}.
This is a rather involved function shown explicitly in \eqref{b}.
The final result for the contribution of the graph (d) is
\beq\label{DES}
\chi_d(z,z')=\chi_0(z{,}z')\;{24\over 5}\,u_0\,(-{\cal P}_D)\,z'^\varepsilon\;
\left[{\zeta^\varepsilon K \over 5+\varepsilon}+{L-\zeta^\varepsilon K \over \varepsilon}+{L\over 5-\varepsilon}+H_d(\zeta)\right]
\eeq
where
\beq\label{dfd}
H_d(\zeta)=f_0(\zeta)+{25 f_1(\zeta)\over 5+\varepsilon}
\left[(1-\zeta)^{3+\varepsilon}+(1+\zeta)^{3+\varepsilon}\right]+
{25 f_2(\zeta)\over 5+\varepsilon}\left[(1-\zeta)^{3+\varepsilon}-(1+\zeta)^{3+\varepsilon}\right]
\eeq
is the new function, absent in the calculations of \cite{ES94}.
It is responsible for the dependence of $\chi_d(z,z')$
(as well as of the entire layer susceptibility)
on arbitrary values of $\zeta$ in the whole interval $(0,1)$.
The somewhat bulky expressions for the functions
$f_0(\zeta)$, $f_1(\zeta)$ and $f_2(\zeta)$ can be found in the Appendix
along with some more calculational details.

The function $H_d(\zeta)$ is regular at $\zeta\to0$ and behaves $\sim\zeta^4$
as predicted in \cite{ES94}:
\beq\nonumber
H_d(\zeta)={1\over 20736}(2-\varepsilon)(6-\varepsilon)(11-\varepsilon)\,\zeta^4+O(\zeta^6).
\eeq
It is also regular as $\varepsilon\to0$, but contains singularities at $\zeta=\pm1$.
The singularity at $\zeta=1$ arises on the mutual approach of coordinates $z$ and $z'$.
It is related to the short-distance behavior inside the bulk of the system,
given by the OPE. We shall return to this issue below.
The value $\zeta=-1$ is formally outside the definition range of $\zeta$, but it
corresponds to a mirror image with respect to the boundary, the configuration
when one of the coordinates $z$ or $z'$ changes its sign.

The first three terms in square brackets of \eqref{DES}
have already been present in \cite[(4.10)]{ES94}.
They contain the leading contribution to $\chi_d(z,z')$ in
the limit $\zeta\to0$ responsible for the short-distance BOE singularity \cite{ES94}.
The epsilon expansion
\beq\label{lns}
{\zeta^\varepsilon K \over 5+\varepsilon}+{L-\zeta^\varepsilon K \over \varepsilon}+{L\over 5-\varepsilon}=
\Big(\frac15-\frac12\ln\zeta\Big)\,\frac1\varepsilon+O(1)
\eeq
shows up a logarithmic behavior in $\zeta$.
The explicit expressions for the constants
$K$ and $L$ are
\beq\label{k}
K= {(2-\varepsilon) (72-\varepsilon^2)\over 6 \varepsilon (2+\varepsilon) (4+\varepsilon) (6+\varepsilon)}
\qquad\mbox{and}\qquad
L={(2-\varepsilon)(4-\varepsilon)(6-\varepsilon)\over 48\varepsilon(2+\varepsilon)}\,.
\eeq
They have to be multiplied by the factor $2^{1+\varepsilon}\Gamma(2-\varepsilon)=2+O(\varepsilon)$
in order to match their counterparts from \cite{ES94} (see Appendix).
In this way $L$ conforms with ${\cal L}$ from \cite[(4.11)]{ES94}.
The constant ${\cal K}$ has been calculated in \cite{ES94} only to order
$O(\varepsilon)$, and ({\ref{k}}) shows it explicitly.

\subsection{The one-loop result and $\varepsilon$ expansion}\label{OCHI}

The complete expression for the layer susceptibility $\chi(z,z')$, which is the sum of
contributions (\ref{chi0}), (\ref{b+c}) and (\ref{DES}), is
\beq\label{chifull}
\chi(z,z')=\frac15 {z^3\over z'^2}
+ {6\over 25}\,u_0\,{S_d^{-1}\over d{-}2}\;2^\varepsilon
\;{z^3\over z'^{2-\varepsilon}}\;
\left[{K{+}S\over 5+\varepsilon}\,\zeta^\varepsilon+{(L{+}S){-}\zeta^\varepsilon (K{+}S) \over \varepsilon}
+{L{+}S\over 5-\varepsilon}+H_d(\zeta)\right].
\eeq
The constant $S$ stems from (\ref{b+c}) and is given by (cf. \cite[(4.8)]{ES94})
\be\nonumber
S=-\,{(4{-}\varepsilon)(6{-}\varepsilon)\over 48\varepsilon(2{+}\varepsilon)}\left[{12\over (1{+}\varepsilon)(4{-}\varepsilon)} -1\right],
\ee
the function $H_d(\zeta)$ is defined in \eqref{dfd} and
$f_i(\zeta)$ are listed in \eqref{0412}-\eqref{2412}.
Again, the first three terms in square brackets of (\ref{chifull})
agree with those of \cite[(4.6), (4.10)]{ES94}. As $\varepsilon\to0$, we have
\beq\label{INS}
{K+S\over 5+\varepsilon}\,\zeta^\varepsilon+{(L+S)-\zeta^\varepsilon (K+S)\over \varepsilon}+{L+S\over 5-\varepsilon}=
{25\over 288}-{5\over 16}\ln\zeta +O(\varepsilon).
\eeq

In the $\varepsilon$ expansion of $\chi(z,z')$ in (\ref{chifull})
we use the standard coupling-constant renormalization
of the massless theory and the corresponding value of the one-loop fixed point.
This implies the chain of transformations
\beq\label{u*}
u_0\to \bar u_0=u_0 \mu^{-\varepsilon} \to u=\bar u_0 K_d \to u^*={2\over 3}\,\varepsilon
\eeq
where $\mu$ is an arbitrary momentum scale and $K_d=(2\pi)^{-d}S_d$.
The $S_d$ can be found in \eqref{Sd}.

Factoring out the overall constant
amplitude and exponentiating the $\ln\zeta$ term from \eqref{INS} we obtain the scaling form
\beq\label{chieps}
\chi(z,z')=\sqrt{4z z'}\;\zeta^{\frac{5-\varepsilon}2}\frac1{10}\left(1+{5\varepsilon\over 36}\right)
\Big[1+\varepsilon\,g(\zeta)\Big]+O(\varepsilon^2)\;.
\eeq
The function $g(\zeta)$ stems directly from $H_d(\zeta)$ in \eqref{dfd} and is given by
\beq\label{gzeta}
g(\zeta)=\frac{\zeta^{-6}}{1260}
\Big[g_0(\zeta )-g_1(\zeta ) (1+\zeta )^3 \ln(1 {+} \zeta )
- g_1(-\zeta ) (1-\zeta )^3 \ln(1 {-} \zeta )\Big]
\eeq
with
\begin{align}\nonumber
&g_0(\zeta )=2700 \zeta^2-1030 \zeta^4+3343 \zeta^6\;, \\
\nonumber
&g_1(\zeta )=30\,\big(8+25 \zeta-36 \zeta^2+25 \zeta^3+8 \zeta^4\big)\,.
\end{align}
It is an even function of its argument and starts, as it should be, with
\be\nonumber
g(\zeta)=\frac{11}{1080}\;\zeta^4+O(\zeta^6)
\ee
as $\zeta\to0$.

\subsection{Scaling}
The explicit expression \eqref{chieps} matches the "improved" scaling form
(cf. \eqref{xisc})
\beq\label{SX}
\chi(z,z')=(4zz')^\frac{1-\eta}{2}\;\zeta^\frac{\eta_\parallel-1}{2}\;Y(\zeta)
\eeq
where the generally expected singular behavior of $\chi(z,z')$ in the limit $\zeta\to0$ is taken
into account (see \cite[(8)]{RJ82a}, \cite[(4.68)]{Gompper86}, \cite[Sec. III.B]{KlD99}).
The power of $\zeta$ involves
the critical exponent of parallel correlations at $T=T_c$, $\eta_\parallel$.
It is related to the (full) scaling dimension $\hat\Delta$
of the leading BOE operator $\hat\phi$ via
$\hat\Delta\equiv\hat\Delta_{\hat\phi}=\frac12(d-2+\eta_\parallel)$
(see e. g. \cite[p. 12]{BHS19}).
The function $Y(\zeta)$ is regular when $\zeta\to0$.

At the extraordinary transition,
the exact value of the correlation exponent $\eta_\parallel$ is $\eta_\parallel=d+2$
\cite[Sec. 5.3]{BM77}, \cite[Sec. III.C.15]{Die86a},
which corresponds to the leading boundary operator with scaling dimension $d$, the
stress-energy tensor $T_{zz}$ \cite{Car90}.
Thus we can write $\chi(z,z')$ in \eqref{chieps} as
\beq\label{chied}
\chi(z,z')=(4zz')^\frac{1-\eta}{2}\;\zeta^{\frac{d+1}2}\frac1{10}\left(1+{5\varepsilon\over 36}\right)
\Big[1+\varepsilon\,g(\zeta)\Big]+O(\varepsilon^2)
\eeq
where the full correlation exponents $\eta$ and $\eta_\parallel$ appear.
No further higher powers of $\varepsilon$ should contribute to the factor $\zeta^{(d+1)/2}$.
The last form of $\chi(z,z')$ could be used for eventual extrapolations to $d=3$ dimensions.

\subsection{The singular behavior at $\zeta\to1$}

In the alternative limit $\zeta\to1$,
the singular behavior of $\chi(z,z')$ is related to the term
$\sim\ln(1-\zeta)$ in (\ref{gzeta}).
The coordinate $z$ approaches $z'$ and the thickness of the layer becomes small.
The non-analyticity associates, via OPE, with the singular behavior of the energy density.
The limit $\zeta\to1$ was unaccessible in \cite{ES94} where calculations have been done
only in the asymptotic regime $\zeta\ll 1$.

To exponentiate of the $\ln(1-\zeta)$ term we
express \eqref{chieps} in a form manifestly symmetric
with respect to interchange of $z$ and $z'$. Thus we write
\be\nonumber
\zeta={z+z'-|z-z'|\over z+z'+|z-z'|} \equiv {1-x\over 1+x}
\qquad\mbox{with}\qquad x=\frac{|z-z'|}{z+z'}\;.
\ee
Expanding $\chi(z,z')$ in powers of $x$ and exponentiating the $\ln x$ term we obtain
\beq\label{altchi}
\chi(z{\leftrightarrow} z')=\frac{z+z'}{10}\left[
a_0-5\,x+a_2\,x^2 +a_3\,x^{3-{2\over 3}\varepsilon}+ O(x^4,\varepsilon^2)\right]
\eeq
with coefficients
\begin{align}\nonumber
&a_0=1+{\varepsilon\over 7} \left( {1297\over 45}- 40\,\ln2 \right)+O(\varepsilon^2)\,,\\
\nonumber
&a_2=12 \left[1 +{\varepsilon\over 21} \left( {3593\over 60}- 86\,\ln{2}\right)\right]+O(\varepsilon^2)\,,\\
\nonumber
&a_3=-20 \left[1-{\varepsilon\over 36}(29+24 \,\ln{2} )\right]+O(\varepsilon^2)\,.
\end{align}

The non-analytic power-law contribution $\sim|z-z'|^{3-2\varepsilon/3}$
can be understood from an OPE argument.
When the distance $|x-x'|$ between points $x$ and $x'$ goes to zero, the OPE of
the product $\phi(x)\phi(x')$ has a contribution proportional to the energy-density operator
$\epsilon({\bar x})=-\frac12\phi^2({\bar x})$,
\beq\label{CF2}
C_\mathrm{\phi\phi}^{\phi^2}(|x-x'|)\,\phi^2(\bar x)\,.
\eeq
The expansion point $\bar x$ is usually defined to be
the midpoint $\bar x=(x+x')/2$. In the semi-infinite geometry,
the one-point function of the energy density has a non-vanishing profile at the transition
point (for a very careful discussion of this and similar profiles see \cite{EKD93,KED95,EKD96})
\beq\label{Fprof}
\la\phi^2({\bar x})\ra=\frac{A_{\phi^2}}{(2\bar z)^{\Delta_{\phi^2}}}\,.
\eeq
Owing to the translational invariance in parallel directions,
only the dependence on the normal distance to the boundary appears on the right-hand side.
The scaling dimension $\Delta_{\phi^2}$ is related to the usual bulk critical exponents
of the specific heat and correlation length via
\be\nonumber
\Delta_{\phi^2}=\frac{1-\alpha}{\nu}=d-\frac1\nu\,.
\ee
By \eqref{CF2}--\eqref{Fprof}, the coordinate dependence of the
short-distance coefficient $C_\mathrm{\phi\phi}^{\phi^2}(|x-x'|)$ is proportional to
$|x-x'|^{-2\Delta_\phi+\Delta_{\phi^2}}$.
Thus the scaling function
$g(\xi)$ of the two-point function \eqref{scalg} has a contribution
$\sim\xi^{-\Delta_\phi+\Delta_{\phi^2}/2}$ as $\xi\to0$.
Integrated over $d-1$ parallel directions, this leads to the singular contribution
\be\nonumber
\chi_{sing}(z{\leftrightarrow}z')\sim|z-z'|^{d-1-2\Delta_\phi+\Delta_{\phi^2}}=
|z-z'|^{1-\eta+\Delta_{\phi^2}}
\ee
in the layer susceptibility.
In the special case of the one-component scalar field considered throughout the paper,
$\Delta_{\phi^2}=2-2\varepsilon/3+O(\varepsilon^2)$, and the exponent
of the last power is just $3-2\varepsilon/3+O(\varepsilon^2)$ in agreement with \eqref{altchi}.

\subsection{A cool-down: Ordinary transition}\label{chiorS}
Within the same one-loop order, at the ordinary transition
$\chi(z,z')$ is given only by two graphs (a) and (c) from (\ref{graphs}).
These are is easy to calculate, and the result is (again, with $z<z'$)
\be\nonumber
\chi^{ORD}(z,z')=z-{u_0\over 2}\,{\cal P}_D z \,z'^\varepsilon
\left({\zeta^\varepsilon \over 1{+}\varepsilon}+{1{-}\zeta^\varepsilon \over \varepsilon}+{1\over 1{-}\varepsilon} \right).
\ee
Using ${\cal P}_D$ from (\ref{PD}), at the fixed point (\ref{u*}) we obtain the $\varepsilon$ expansion
\beq\label{chiord}
\chi^{ORD}(z,z')=z\Big[1+{\varepsilon\over 6}(2-\ln\zeta)\Big]+O(\varepsilon^2)
=\frac12\Big(1+{\varepsilon\over 3}\Big) \sqrt{4z z'}\;
\zeta^{{1{-}\varepsilon/3\over 2}}+O(\varepsilon^2)\,.
\eeq

The power of $\zeta$ here agrees again with the expected one from \eqref{SX}.
Indeed, now $\eta_\parallel=2-\varepsilon/3+O(\varepsilon^2)$ following, e. g.,  from
\cite[(3.155a)]{Die86a} with $n=1$. Hence $(\eta_\parallel-1)/2=(1-\varepsilon/3)/2+O(\varepsilon^2)$
in agreement with \eqref{chiord}.
A similar calculation could be easily done for the special transition.
In this case $\eta_\parallel^{sp}=-\varepsilon/3+O(\varepsilon^2)$ \cite[(3.156a)]{Die86a}.

In \eqref{chiord}, there is no correction of order $O(\varepsilon)$ to the pure power-law behavior
in $\zeta$ as in (\ref{chieps}).
The origin of that correction was the
contribution of the graph (d) appearing in the presence of a non-vanishing
order-parameter profile at the extraordinary transition.
Accordingly, the singular behavior of $\chi(z,z')$ as $\zeta\to 1$ could be viewed
as a consequence of the mutual approach of two $\phi^3$ vertices in the graph (d).
For ordinary transition, an analogous correction is expected at
order $O(\varepsilon^2)$ from the two-loop sunset graph.

\section{Discussion and outlook}\label{D:O}

In 1995, McAvity and Osborn applied the integral transformations \cite[(4.18)-(4.19)]{MO95} in their prominent investigation of
"Conformal field theories near a boundary in general dimensions"
in the context of a large-$N$ expansion.
They acknowledged an analogy with the Radon transformation.
In doing so, they referred to the book \cite{GGV66}.
Some other mathematical references on the subject can be mentioned \cite{Ludwig66,Deans83,Helgason99}.
The statement was that by integrating \eqref{scalg} over planes parallel to the boundary, the transform function \eqref{gro} is obtained, which can be subsequently inverted to reproduce \eqref{scalg}.
Apart from the application of this procedure in \cite[Sec. 4]{MO95} and its subsequent discussion in \cite{McA95}, we did not find any other instance of its use.

In the present paper, we have presented a successful
application of the Radon transform, as it was formulated in \cite[Sec. 4]{MO95},
to the mean-field propagator at the extraordinary transition.
This allowed to obtain its real-space representation, continuously dependent on
space dimension $d$, on the zeroth loop level of the Landau-Ginzburg theory.
Its simple and symmetric form enabled further explicit calculations of the parallel susceptibility $\chi(z,z')$ in the one-loop approximation and its $\varepsilon=4-d$ expansion
to order $O(\varepsilon)$.

Actually, an inverse transform of $\chi(z,z')$ from Sec. \ref{OCHI}
produces the two-point correlation function at the extraordinary transition to $O(\varepsilon)$
\cite{DHS20}, which has never been calculated before.
As a simple exercise, the inversion formula could be applied to $\chi^{ORD}(z,z')$ from Sec. \ref{chiorS} to reproduce the known $O(\varepsilon)$ expression for the two-point correlator
at the ordinary transition \cite{GW85,MO93,MO95,BHS19}. Its extension to the $O(\varepsilon^2)$ order 
must be straightforward.

Recently, in this less complicated case, the two-point function has been analytically obtained to $O(\varepsilon^2)$ from the boundary conformal bootstrap \cite{BHS19}.
There was no such explicit result available for this function:
only overcomplicated momentum  \cite[(B.45)-(B.46)]{DD81c}
and integral \cite[(16)]{OO85} representations existed before
in the statistical-physics literature.
Hence, it would be useful to derive it by other means and to perform the appropriate comparison.
An application of the present approach should be capable of reaching this goal.
That would be useful both for the statistical mechanics and boundary conformal field theory.

Applications to different models and in other fields beyond the statistical physics
would be of interest.

\vspace{5mm}
\noindent{\bf Acknowledgments.}\ A part of this work has been carried out at
the Fakult\"at f\"ur Physik of the Universit\"at Duisburg-Essen in 1995.
The author is grateful to
H. W. Diehl for suggesting the problem, his warm hospitality and financial support,
as well as for enlightening correspondence in preparation of the final version of the paper.
Collaboration with H. W. Diehl and A. Drewitz in the initial stage of this work is
gratefully acknowledged. The author is indebted to H. Osborn for bringing Ref. \cite{LRR13} to his attention, and to G. Gompper
for sending him a hard copy of his PhD Thesis \cite{Gompper86}.
The author thanks to the anonymous referee for bringing to his attention Refs.
\cite{Chester20} and \cite{Lewellen}.
\vspace{4mm}

\appendix
\section{Details related to the graph (d)}

Here we give some technical details of our calculation of the "hard" graph (d)
and list the explicit expressions for the functions $f_0(\zeta)$, $f_1(\zeta)$ and $f_2(\zeta)$
appearing in the function $H_d(\zeta)$ in \eqref{dfd}.

For an explicit calculation of the contribution $\chi_d(z,z')$
in (\ref{d}) we need an exact expression for the inner loop of the graph (d),
\be\nonumber
{\cal B}(y,y')=\int\!\! d^{d-1}r\,\,G_0^2(r;y,y')\equiv C_d\;2^\varepsilon\,b(y,y')\;.
\ee
With the function $G_0(r;y,y')$ from (\ref{prop}), a straightforward calculation yields
\begin{align}\label{b}
b&(y,y'|y{<}y')={1\over 4}{1\over 1-\varepsilon}\Big(|y_-|^{-1+\varepsilon}+y_+^{-1+\varepsilon} \Big)
+{12-\varepsilon\over 2 \varepsilon (1-\varepsilon)}\; y'^{-1+\varepsilon} \\
&- {3\over \varepsilon (1+\varepsilon)}\;{1\over y y'} \Big(|y_-|^{1+\varepsilon}-y_+^{1+\varepsilon} \Big)
- 3 {6+5\varepsilon\over \varepsilon (1+\varepsilon)(2+\varepsilon)(3+\varepsilon)}\;{1\over y^2 y'^2}
\Big(|y_-|^{3+\varepsilon}+y_+^{3+\varepsilon} \Big)\nonumber\\
&+{6\over \varepsilon (1-\varepsilon^2)(2+\varepsilon)(3+\varepsilon)}\;{1\over y^2}
\Big[4(3-5\varepsilon-\varepsilon^2) \,y'^{1+\varepsilon}-(2-5\varepsilon)(3+\varepsilon)\,|y_-|\,y_+\;y'^{-1+\varepsilon}
\Big] \nonumber\\
&- {36\over \varepsilon (2+\varepsilon)(3+\varepsilon)(5+\varepsilon)}\;{1\over y^3 y'^3}
\left[|y_-|^{5+\varepsilon}-y_+^{5+\varepsilon}+{1\over 7+\varepsilon} \;{1\over yy'}
\Big(|y_-|^{7+\varepsilon}+y_+^{7+\varepsilon}\Big)\right]\nonumber\\
&- {72\over \varepsilon (1-\varepsilon^2)(2+\varepsilon)(3+\varepsilon)(5+\varepsilon)}
\;{1\over y^4}\Big[{48\over 7+\varepsilon}\; y'^{3+\varepsilon}-12\,|y_-|\,y_+\;y'^{1+\varepsilon}+
(5+\varepsilon)\,|y_-|^2\,y_+^2\; y'^{-1+\varepsilon}\Big]. \nonumber
\end{align}
Here we use the short-hand notation $y_-=y'-y$ and $y_+=y+y'$.

It is important that due to the full symmetry
$b(y,y')=b(y',y)$, the double integration in (\ref{d})
can be reduced to a single one. This has been done in \cite[(4.9)]{ES94}.
Using this single-integral representation and (\ref{b}), we obtained
(\ref{DES})-(\ref{dfd}).
Note that the function $g({\cal Z})$ in \cite{ES94}
is related to $b(y'/y){\equiv} b(1,y'/y)$ through the normalization
\be\nonumber
g\Big({y'\over y}\Big)=2^{1+\varepsilon}\Gamma(2-\varepsilon)\,y^{1-\varepsilon}\,b(y,y'|y{<}y')\equiv
2^{1+\varepsilon}\Gamma(2-\varepsilon)\,b\Big({y'\over y}\Big)\quad\mbox{for}\quad{y'\over y}>1\,.
\ee
This means that all integrals involving the function $b$
and, in particular, constants $K$ and $L$ (see (\ref{DES}) and below) have to be
multiplied by $2^{1+\varepsilon}\Gamma(2{-}\varepsilon){=}2{+}O(\varepsilon)$ to
retrieve their counterparts from \cite{ES94}.

\vskip3mm
The functions $f_i(\zeta)$ from \eqref{dfd} with $\zeta=z/z'\le 1$, are
\begin{align}
&f_0(\zeta)={5 (2-\varepsilon) (7-\varepsilon)\over 48 \varepsilon (1-\varepsilon^2)}
- {25 (5-\varepsilon)\; \zeta^{-2}\over 12 (1-\varepsilon^2) (2+\varepsilon) (3+\varepsilon)}+
 {75 (3-\varepsilon)\; \zeta^{-4}\over 2 \varepsilon (1-\varepsilon^2) (3+\varepsilon) (4+\varepsilon) (5+\varepsilon)}
\nonumber\\
\label{0412}
&\quad\qquad+{300\; \zeta^{-6}\over \varepsilon (1+\varepsilon) (2+\varepsilon) (3+\varepsilon) (5+\varepsilon) (6+\varepsilon) (7+\varepsilon)}\;,
\\\label{1412}
&f_1(\zeta)=-{24 (1-\varepsilon) (4+\varepsilon) (\,\zeta^{-2}+\zeta^{-6})
-(432+282 \varepsilon+67 \varepsilon^2+2 \varepsilon^3+\varepsilon^4)\; \zeta^{-4} \over
4 \varepsilon (1-\varepsilon^2) (2+\varepsilon) (3+\varepsilon) (4+\varepsilon) (6+\varepsilon) (7+\varepsilon)}\;,
\\
\label{2412}
&f_2(\zeta)={(50+\varepsilon+5 \varepsilon^2) (\,\zeta^{-3}+\zeta^{-5})\over
2 \varepsilon (1-\varepsilon^2) (2+\varepsilon) (4+\varepsilon) (6+\varepsilon) (7+\varepsilon)} \;.
\end{align}

\providecommand{\href}[2]{#2}\begingroup\raggedright\endgroup

\end{document}